\documentstyle[epsfig,12pt]{article}
\def \bea{\begin{eqnarray}}
\def \beq{\begin{equation}}
\def \bo{B^0}

\def \eea{\end{eqnarray}}
\def \eeq{\end{equation}}

\def \hp{\hat{p}}

\def \mat#1#2{\langle #1 | #2 \rangle}
\def \ob{\overline{B}^0}
\def \od{\overline{D}^0}

\def \pr{\parallel}

\begin{document}
\Large
\centerline {\bf Extraction of a Weak Phase from $B \to D^{(*)} \pi$
\footnote{ANL-HEP-PR-01-086, Enrico Fermi Institute preprint EFI 01-35,
hep-ph/0110159. To be submitted to Physical Review D.}}
\normalsize
\bigskip
 
\centerline {Denis A. Suprun$\,^a$~\footnote{d-suprun@uchicago.edu}, Cheng-Wei
Chiang$\,^{a,b}$~\footnote{chengwei@hep.uchicago.edu}  and
Jonathan L. Rosner$\,^a$~\footnote{rosner@hep.uchicago.edu}}
\vspace{0.5cm}
\centerline{\it $^a$ Enrico Fermi Institute and Department of Physics}
\centerline{\it University of Chicago, 5640 S. Ellis Avenue, Chicago, IL 60637}
\vspace{0.2cm}
\centerline{\it $^b$ HEP Division, Argonne National Laboratory}
\centerline{\it 9700 S. Cass Avenue, Argonne, IL 60439}
\bigskip
 
\begin{quote}

To observe CP-violating asymmetries through the interference of a weaker
amplitude with a stronger one in $\bo \to D^{(*)} \pi$ and $\ob \to D^{(*)}
\pi$ decays, one must collect enough events that the intensity
associated with the weaker amplitude would be statistically
significant. We show that provided the weaker amplitude is measured separately
in $B^\pm \to D^{(*)\pm} \pi^0$ decays, the time-integrated approach requires
around $2.5 \cdot 10^8 \ B \bar B$ pairs for measurements of the weak phase
$\sin(2\beta + \gamma)$ with an uncertainty of $0.05$ or better. We also
determine the optimal conditions for precise $2\beta + \gamma$ measurements
and discuss the possibilities for resolving a discrete ambiguity. 

\end{quote}

\noindent
PACS Categories:  13.25.Hw, 14.40.Nd, 14.65.Fy, 12.15.Hh

\section{Introduction}

The phases of elements of the Cabibbo-Kobayashi-Maskawa (CKM) matrix
describing the weak charge-changing interactions of quarks are of fundamental
importance.  Together with magnitudes of the matrix elements and masses of the
six quarks $(u,c,t)$ and $(d,s,b)$, these phases must be explained by any
theory which extends our knowledge beyond the Standard Model (SM) of
electroweak and strong interactions.

Indirect information on CKM phases \cite{CKMrevs,JRTASI} is now being
supplemented by measurements of CP-violating asymmetries in $B$ decays
\cite{earlybeta,sin2beta,sin2betaBelle} which provide direct phase
measurements.
The weak phase $\beta \equiv {\rm Arg}(-V^*_{cb}V_{cd}/V^*_{tb}V_{td})$ is
determined by measurements of the rate asymmetry in decays such as $B^0 \to
J/\psi K_S$, while $\alpha \equiv {\rm Arg}(-V^*_{tb}V_{td}/V^*_{ub}V_{ud})$
will be determined by measurements in decays such as $B \to \pi \pi$ and $B
\to \rho \pi$.  Information on all charge modes will be needed to separate
contributing amplitudes from one another
\cite{GrLSQS}.

Information on $\gamma \equiv {\rm Arg}(-V^*_{ub}V_{ud}/V^*_{cb}V_{cd})$ is
more difficult to obtain.  The decays $B^\pm \to D^0 K^\pm$, $B^\pm \to \od
K^\pm$, and $B^\pm \to D_{CP} K^\pm$, where $D_{CP}$ is a CP eigenstate, permit
one to perform a triangle construction to extract the weak phase $\gamma$
\cite{GW}.  The interference of the Cabibbo-favored decay $D^0 \to K^- \pi^+$
and the doubly-Cabibbo-suppressed decay $D^0 \to K^+ \pi^-$ introduces an
important subtlety in this method \cite{but}. Numerous determinations of
$\gamma$ using nonstrange and strange $B$ decays to $\pi \pi$ and $K \pi$ are
subject to questions associated with SU(3) flavor violation, electroweak
penguin contributions, and rescattering \cite{JRTASI}.

The Cabibbo-favored decays $B^0 \to D^{(*)-} \pi^+$ and $\ob \to D^{(*)+}
\pi^-$ and the corresponding doubly-Cabibbo-suppressed modes $\ob \to D^{(*)-}
\pi^+$ and $B^0 \to D^{(*)+}K \pi^-$ can provide information on the weak phase
$2 \beta +\gamma$ \cite{DR,DSetal,D98,ASY}. (One can substitute 
$\rho^\pm$ or $a_1^\pm$ for the charged pion.) These methods typically
require measuring either a very small rate asymmetry (for the Cabibbo-favored
modes) or a very small rate (for the Cabibbo-suppressed modes).  It was
therefore suggested recently \cite{LSS} that one instead measure $2 \beta +
\gamma$ via the interference of a small amplitude with a larger one in
decays of the form $B \to V_1 V_2$, where, for example, $V_1 = D^*$ and
$V_2 = \rho$.  The interference is to be detected through characteristic
angular distributions in decay products of the vector mesons, and through
time-dependent measurements.  Refs.\ \cite{ASY} and \cite{CW} contain some
useful results regarding these distributions.

In the present paper we analyze the possibilities of precise measurements of
$2\beta + \gamma$ for the simplest case of $B \to D^{(*)} \pi$ decays.
We find the optimal conditions for measuring $2\beta + \gamma$.  We also
estimate the number of $B \bar B$ pairs needed for such measurements that 
will reduce the allowed range of $2\beta + \gamma$ values to the currently
achieved indirect bounds coming from measurements of other CKM parameters. 

A general feature of CP-violating asymmetries detected through the
interference of a weaker amplitude with a stronger one is that one must be able
to detect processes at the level of the {\it absolute square of the weaker
amplitude} \cite{EGR}.  We find that this situation holds for $B \to D^{(*)}
\pi$ decays.  One still has to be able to collect enough events such that the
absolute square of the Cabibbo-suppressed amplitude would be detectable with
good statistical significance.  This translates to the need for several times
$10^8$ produced $B \bar B$ pairs. (Ref.~\cite{LSS} cites a figure of $10^8$
pairs for a useful measurement of $\sin(2 \beta + \gamma)$ using $B \to V_1
V_2$ decays.)  In fact, our best determination makes use of a direct
measurement of the weaker amplitude through a factorization relation between
$B^0 \to D^{(*)+} \pi^-$ and $B^+ \to D^{(*)+} \pi^0$ \cite{D98}.
For both pseudoscalar and vector $D$ mesons in the final state, we employ 
different models to anticipate the size of the weaker amplitude. However, 
direct measurements of the rates for $B^+ \to D^+ \pi^0$ and $B^+ \to D^{*+} 
\pi^0$ will eventually give us these amplitudes directly.

In Section II we introduce our notation and predictions for decay rates of
neutral $B$ mesons in the framework of the time-integrated approach. We shall
quote results for $B \to D^* \pi$ decays because of advantages in $D^*$
detection, recognizing that many are also valid for $B \to D \pi$.
Decay rates as functions of a minimum vertex separation (expressed in terms of
proper time) are of particular interest to us in Section III as we try to find
the optimal conditions for measuring the weak phase $2\beta + \gamma$ with high
precision. In Section IV we circumvent the problem of measuring the small
weaker-to-stronger amplitude ratio $R$ by making a foray into charged $B$ meson
decays, using the process $B^+ \to D^{(*)+} \pi^0$. Estimates of the
minimum number of $B \bar B$ pairs required for precise measurements of $2\beta
+ \gamma$ are obtained in Section V.  These are convoluted with a finite
time-resolution function and realistic mistagging probabilities in Section VI.
In Section VII we discuss a possibility of partial resolution of an 8-fold
discrete ambiguity by separating $2\beta + \gamma$ and the strong phase
$\delta$ between Cabibbo-allowed and Cabibbo-suppressed modes. We summarize
our results in Section VIII.  

\section{Notation and predictions}

The ``right-sign'' decays $\bo \to D^{*-} \pi^+$ and $\ob \to D^{*+} \pi^-$ 
are governed by the Cabibbo-favored combination of CKM matrix elements 
$V^*_{cb}V_{ud}$ or charge-conjugate, while the ``wrong-sign'' decays $\ob \to
D^{*-} \pi^+$ and $\bo \to D^{*+} \pi^-$ are governed by the 
doubly-Cabibbo-suppressed combination $V^*_{cd}V_{ub}$ or charge-conjugate.  
We denote $f \equiv D^{*-} \pi^+$ and $\bar f \equiv D^{*+} \pi^-$. Then from
\beq
\mat{f}{\bo} = A_1 e^{i \phi_1} e^{i \delta_1}~~,~~~
\mat{f}{\ob} = A_2 e^{i \phi_2} e^{i \delta_2}
\eeq
it follows that
\beq
\mat{\bar f}{\ob} = A_1 e^{-i \phi_1} e^{i \delta_1}~~,~~~
\mat{\bar f}{\bo} = A_2 e^{-i \phi_2} e^{i \delta_2}~~,
\eeq
where the weak phases $\phi_i$ change sign under CP conjugation, while the
strong phases $\delta_i$ do not.  The amplitudes are in the ratio
\beq
\label{def:r}
R \equiv \frac{A_2}{A_1} = \left| \frac{V^*_{cd}V_{ub}}{V^*_{cb}V_{ud}}
\right| r
= |- \lambda^2 (\rho - i \eta)| r \simeq 0.02 r~~,
\eeq
where $\lambda \simeq 0.22$, $\rho$, and $\eta$ are parameters \cite{WP}
which describe CKM matrix elements, and $r = {\cal O}(1)$ describes a
ratio of decay constants and form factors. The weak phase difference is
\beq
\phi_1 - \phi_2 = {\rm Arg} \left( \frac{V^*_{cb}V_{ud}}{V^*_{cd}V_{ub}}
\right) = \pi + \gamma~~.
\eeq

We write the time-dependent decay amplitudes in terms of the functions
\cite{DR} 
\beq
f_+(t) \equiv e^{- i m t} e^{- \Gamma t/2} \cos(\Delta m t/2)~~,~~~
f_-(t) \equiv e^{- i m t} e^{- \Gamma t/2} i \sin(\Delta m t/2)~~,
\eeq
where $m = (m_L + m_H)/2$ is the average of the two mass eigenvalues, $\Delta
m = m_H - m_L$ is their difference, $\Gamma = (\Gamma_1 + \Gamma_2)/2$ is
the average decay rate of the eigenstates, and we neglect $\Delta \Gamma =
\Gamma_H - \Gamma_L$.  Then
\bea
\mat{f}{\bo(t)} & = & f_+(t) \mat{f}{\bo} + \frac{q}{p}f_-(t) \mat{f}{\ob}
\nonumber \\
& = & e^{-imt} e^{-\Gamma t/2} \left( A_1 e^{i \phi_1} e^{i \delta_1}
\cos \frac{\Delta m t}{2} + i \frac{q}{p} A_2 e^{i \phi_2} e^{i \delta_2}
\sin \frac{\Delta m t}{2} \right)~~, \\
\mat{\bar f}{\ob(t)} & = &f_+(t) \mat{\bar f}{\ob} + \frac{p}{q} f_-(t)
\mat{\bar f}{\bo} \nonumber \\
& = & e^{-imt} e^{-\Gamma t/2} \left( A_1 e^{-i \phi_1} e^{i \delta_1}
\cos \frac{\Delta m t}{2} + i \frac{p}{q} A_2 e^{-i \phi_2} e^{i \delta_2}
\sin \frac{\Delta m t}{2} \right)~~.
\eea
If $\bo$--$\ob$ mixing is described primarily by standard model loop
contributions dominated by intermediate $t \bar t$ pairs, we have $q/p =
e^{-2 i \beta}$, and
\bea
\left| \mat{f}{\bo(t)} \right|^2 & = & 
\frac{A_1^2}{2} e^{-\Gamma t} \left[1 + R^2 + ( 1 - R^2 ) \cos \Delta m t 
\right. \nonumber \\
& - & \left. 2R \sin(2 \beta + \gamma - \delta) \sin \Delta m t
 \right]~~, \label{eqn:fbo}\\
\left| \mat{\bar f}{\ob(t)} \right|^2 & = & \frac{A_1^2}{2} e^{-\Gamma t}
\left[1 + R^2 + ( 1 - R^2 ) \cos \Delta m t \right. \nonumber \\
& + & \left. 2R \sin(2 \beta + \gamma + \delta) \sin \Delta m t  \right]~~ 
\label{eqn:bfob},
\eea
where $\delta \equiv \delta_2 - \delta_1$.

Retracing the above steps for the ``wrong-sign''decays $\bo \to D^{*+} \pi^-$
and $\ob \to D^{*-} \pi^+$, we find
\bea
\left| \mat{\bar f}{\bo(t)} \right|^2 & = & \frac{A_1^2}{2} e^{-\Gamma t}
\left[1 + R^2 - ( 1 - R^2 )
\cos \Delta m t \right. \nonumber \\
& - & \left. 2R \sin(2 \beta + \gamma + \delta) \sin \Delta m t
\right]~~,  
\label{eqn:bfbo} \\
\left| \mat{f}{\ob(t)} \right|^2 & = & \frac{A_1^2}{2} e^{-\Gamma t}
\left[1 + R^2 - ( 1 - R^2 ) \cos
\Delta m t \right. \nonumber \\
& + & \left. 2R \sin(2 \beta + \gamma - \delta) \sin \Delta m t
 \right]~~ \label{eqn:fob}.
\eea

Let us now consider the production of a $\bo$$\ob$ pair in $e^+ e^-
\to \Upsilon(4S) \to \bo \ob$, so that the pair is in a state $\Psi_-$ of
negative charge-conjugation eigenvalue.  Assume that we ``tag'' the initial
production of a $\ob(\hp)$ with a $\bo(-\hp)$, and the initial production
of a $\bo(\hp)$ with a $\ob(-\hp)$.  Then if we define the proper decay time
of the state $f$ with center-of-mass direction $\hp$ as $t_f$, that of
the tagging state with direction $-\hp$ as $t_t$, and $t' \equiv t_f - t_t$,
$T \equiv t_f + t_t$, we find \cite{JRTASI,ASY,BaBartd}
\bea
|\mat{\bo(-\hp),D^{*\mp} \pi^\pm(\hp)}{\Psi_-}|^2 & = &
e^{- \Gamma T} |A_1|^2 \left[1 + R^2 \pm (1-R^2) \cos \Delta m t' \right. 
\nonumber \\
& - & \left. 2R \sin(2 \beta + \gamma \mp \delta) \sin \Delta m t' \right]~~,
\nonumber \\
|\mat{\ob(-\hp),D^{*\pm} \pi^\mp(\hp)}{\Psi_-}|^2 & = &
e^{- \Gamma T} |A_1|^2 \left[1 + R^2 \pm (1-R^2) \cos \Delta m t' \right. 
\nonumber \\
& + & \left. 2R \sin(2 \beta + \gamma \pm \delta) \sin \Delta m t' \right]~~.
\eea

One can express the time-integrated decay rates as
\bea
\int_0^\infty dt_f \int_0^\infty dt_t |\mat{\bo(-\hp),D^{*\mp} \pi^\pm(\hp)} 
{\Psi_-}|^2 & \propto & \int_{-\infty}^\infty
dt' e^{- \Gamma|t'|} [A_\pm(t') + B_\mp(t')]~~, \nonumber \\
\int_0^\infty dt_f \int_0^\infty dt_t |\mat{\ob(-\hp),D^{*\pm} \pi^\mp(\hp)}
{\Psi_-}|^2 & \propto & \int_{-\infty}^\infty
dt' e^{- \Gamma|t'|} [A_\pm(t') - B_\pm(t')]~~,
\eea
where
\beq
A_\pm(t') \equiv (1+R^2) \pm (1-R^2) \cos \Delta m t'~~,
\eeq
\beq
B_\pm (t') \equiv - 2 \, R \sin(2 \beta + \gamma \pm \delta) \sin \Delta m t'
\eeq
are even and odd functions of $t'$, respectively.

Now we introduce notation for measurable decay numbers. 
The number of $\bo \to D^{*-} \pi^+$ decays with vertex separation $t' > 0$ is
\beq
N^r_+  \propto  \int_0^\infty dt' e^{- \Gamma |t'|} [A_+(t') + B_-(t')],
\label{eqn:npr}
\eeq        
while those with $t' < 0$ is
\beq
N^r_-  \propto  \int_{-\infty}^0 dt' e^{- \Gamma |t'|} [A_+(t') + B_-(t')] = 
\int_0^\infty dt' e^{- \Gamma |t'|} [A_+(t') - B_-(t')].
\eeq 
Here the superscript ``$r$'' denotes right-sign decays.  The corresponding
expressions $N^w_+$ for the wrong-sign (superscript ``$w$'') decays
$\bo \to D^{*+} \pi^-$ with $t' > 0$ and $N^w_-$ for $\bo \to D^{*+} \pi^-$ 
with $t' <  0$ are
\beq
N^w_\pm \propto  \int_0^\infty dt' e^{- \Gamma |t'|} [ A_-(t') \pm B_+(t')]. 
\eeq
Similar expressions for $\ob$ decays are
\beq
\overline{N^r_\pm} \propto  \int_0^\infty dt' e^{- \Gamma |t'|} [ A_+(t') \mp
B_+(t')], 
\eeq
\beq
\overline{N^w_\pm} \propto  \int_0^\infty dt' e^{- \Gamma |t'|} [ A_-(t') \mp
B_-(t')]. 
\label{eqn:npmwr}
\eeq
Note that the following 4 linear relations among the 8 decay numbers 
\bea
N^{r,w}_+ + N^{r,w}_- = \overline{N^{r,w}_+} + \overline{N^{r,w}_-}, \\
N^{r,w}_+ - N^{r,w}_- = \overline{N^{w,r}_-} - \overline{N^{w,r}_+},
\eea
limit the number of independent quantities to 4. In principle, that allows one
to forgo measurements of $\ob$ decay numbers. However, that method would lead 
to larger uncertainties in determination of $2\beta+\gamma$ and we shall not 
use it.

We shall investigate the dependence of the time-integrated rates on a minimum
vertex separation $t_0$.  The aim of the calculation is to find the optimal
conditions for measuring $\sin(2\beta +\gamma)$.  Fig.~1 shows that
indirect bounds on that weak phase coming from measurements of other CKM
parameters \cite{JRTASI,JRAndr} limit the expected value of $\sin(2 \beta + 
\gamma)$ to the region between $0.89$ and $1$.  To get in the same ballpark 
with the indirect bounds we will calculate the number of $B \bar B$ pairs 
required to determine $\sin(2 \beta + \gamma)$ with an uncertainty of $0.05$. 
This is the main goal of the paper.
\begin{figure}
\centerline{\epsfysize = 4.5 in \epsffile{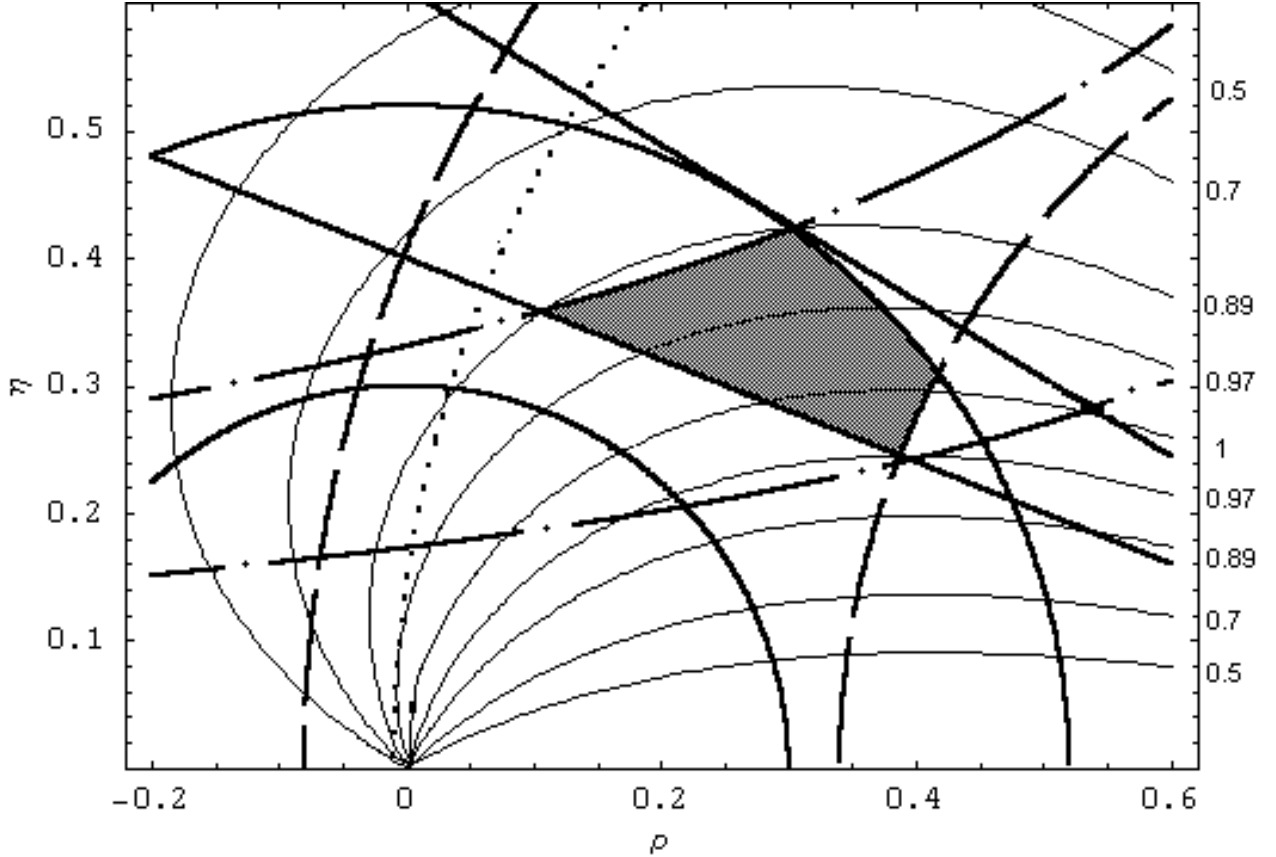}} 
\caption{Contours of $\sin(2\beta+\gamma)$ (thin curves with values to the
right) in ($\rho$,$\eta$) plane.  Thick lines denote current limits on CKM
matrix parameters \cite{JRTASI,JRAndr}.  Solid circles denote limits on
$|V_{ub}/V_{cb}|$ from charmless $b$ decays, dashed circles denote limits on
$V_{td}$ from $\bo-\ob$ mixing, and the dotted circle denotes the lower limit
on $|V_{ts}/V_{td}|$ from the lower limit on $B_s-\overline B_s$ mixing.
Dot-dashed hyperbolae come from limits on CP-violating $K^0-\overline
K^0$ mixing (the parameter $\epsilon$). Two solid rays denote the recent world
average $\pm 1 \sigma$ limits $\sin(2\beta) = 0.79 \pm 0.10$ from neutral $B$
meson decays.  The allowed range is shaded gray.} 
\end{figure}

\section{Decays with vertex separation greater than $t_0$}

If one only takes into account decays with vertex 
separation greater than $t_0$,  Eqs.~(\ref{eqn:npr}--\ref{eqn:npmwr}) become
\beq
N^r_\pm(t_0) \propto \int_{t_0}^\infty dt'\,e^{-\Gamma |t'|} 
[A_+(t') \pm B_-(t')], 
\label{eqn:nr} 
\eeq
\beq
N^w_\pm(t_0) \propto \int_{t_0}^\infty dt'\,e^{-\Gamma |t'|} 
[A_-(t') \pm B_+(t')], 
\eeq
\beq
\overline{N^r_\pm}(t_0) \propto \int_{t_0}^\infty dt'\,e^{-\Gamma |t'|} 
[A_+(t') \mp B_+(t')], 
\eeq
\beq
\overline{N^w_\pm}(t_0) \propto \int_{t_0}^\infty dt'\,e^{-\Gamma |t'|} 
[A_-(t') \mp B_-(t')]. 
\label{eqn:nw}
\eeq
There are several ways to combine these decay numbers together into algebraic 
sums. Some of the resulting combinations may include one of the following 
expressions: $A_+(t') + A_-(t')$, $A_+(t') - A_-(t')$, $B_+(t') + B_-(t')$, 
or $B_+(t') - B_-(t')$. Composing ratios of these algebraic sums (see 
$f_1(t_0)$, $f_2(t_0)$ and $f_3(t_0)$ below), we can extract the parameters 
$R$, $\sin(2\beta+\gamma)\cos\delta$ and $\cos(2\beta+\gamma)\sin\delta$:
\beq
R = \sqrt{\frac{a(t_0)-f_1(t_0)}{a(t_0)+f_1(t_0)}},
\eeq
\beq
SC \equiv \sin(2\beta+\gamma)\cos\delta = \frac{1+R^2}{2b(t_0)\,R}\,f_2(t_0),
\label{eqn:SC}
\eeq
\beq
CS \equiv \cos(2\beta+\gamma)\sin\delta = \frac{1+R^2}{2b(t_0)\,R}\,f_3(t_0),
\label{eqn:CS}
\eeq
where  
\beq
a(t_0) \equiv \Gamma e^{\Gamma t_0} \int_{t_0}^\infty dt'\,e^{-\Gamma t'} 
\cos(\Delta m t') =
\frac{1}{\sqrt{1+x_d^2}}\,\cos(x_d \Gamma t_0 + \Delta), 
\eeq
\beq
b(t_0) \equiv \Gamma e^{\Gamma t_0} \int_{t_0}^\infty dt'\,e^{-\Gamma t'} 
\sin(\Delta m t') =
\frac{1}{\sqrt{1+x_d^2}}\,\sin(x_d \Gamma t_0 + \Delta),
\eeq
\beq
\Delta=\arctan x_d, \qquad x_d \equiv \Delta m/\Gamma,
\eeq
and
\beq
f_1(t_0) \equiv \frac{(N^r_+ + N^r_- + \overline{N^r_+} + \overline{N^r_-}) -
(N^w_+ + N^w_- + \overline{N^w_+} + \overline{N^w_-})}
{N},
\eeq
\beq
f_2(t_0) \equiv \frac{(N^r_- + N^w_- + \overline{N^r_+} + \overline{N^w_+}) -
(N^r_+ + N^w_+ + \overline{N^r_-} + \overline{N^w_-})}
{N},
\label{eqn:f2}
\eeq
\beq
f_3(t_0) \equiv \frac{(N^r_+ + N^w_- + \overline{N^r_+} + \overline{N^w_-}) -
(N^r_- + N^w_+ + \overline{N^r_-} + \overline{N^w_+})}
{N},
\label{eqn:f3}
\eeq
with 
\beq
\label{eqn:totalN}
N \equiv N^r_+ + N^r_- + \overline{N^r_+} + \overline{N^r_-} + N^w_+ + N^w_- 
+ \overline{N^w_+} + \overline{N^w_-}.
\eeq
We have suppressed $(t_0)$ after the decay numbers in the last four formulae.

It has been noted in~\cite{D98,LSS} that $R$ is too small to be
determined by this method. Indeed, calculations show that the smallest 
uncertainty in $R$ is achieved at $t_0=0$ and is equal to 
\beq
\sigma(R)= \sqrt{\frac{x_d^2\,(2+x_d^2)}{16R^2}\,
\frac{1}{\epsilon({\cal B}^r+{\cal B}^w)N_B}}
\approx 0.03,
\eeq
with $\epsilon$ being the tagging efficiency.  We take $\epsilon$ to be 
$0.684
\pm 0.007$~\cite{sin2beta}.  ${\cal B}^r$, the branching ratio of the
``right-sign" decays $B^0 \to D^{*-} \pi^+$, equals $(2.76 \pm 0.21) \times
10^{-3}$~\cite{PDG}. One can show that for $x_d \cong 0.756 \pm 0.012$
\cite{LEPBOSC} the branching ratio of ``wrong-sign" decays is
${\cal B}^w=k{\cal B}^r \approx  0.61 \cdot 10^{-3}$, with 
$k \approx x_d^2/(2+x_d^2) \approx 0.22$.

The error $\sigma(R)=0.03$ is bigger than the approximate $R$ value itself
[Eq.~(\ref{def:r})]. Thus, one has to search for another method of
measuring $R$.

\section{Ratio of amplitudes}

The main reason one cannot get $R$ directly from the ratio of $\bo \to
D^{(*)+} \pi^-$ and $\bo \to D^{(*)-} \pi^+$ decay rates is that the large
$\bo \to \ob$ mixing amplitude in the former overwhelms the smaller 
direct tree contribution.  One can circumvent this obstacle by considering
decays of {\it charged} $B$ mesons, e.g. $B^{\pm} \to D^{(*)\pm} \pi^0$, as
suggested in~\cite{D98}.  The tree amplitude is dominant in these decays and is
proportional to $A_2^2/2$. Thus, the ratio of $B^{\pm} \to D^{(*)\pm} \pi^0$
and $\bo \to D^{(*)-} \pi^+$ decay rates can be used to provide a simple way to
estimate $R$.

The $B^+ \to D^{(*)+} \pi^0$ decay rate can be estimated by assuming
factorization:
\beq {\cal M} = \frac{G_F}{2} V_{ub}^* V_{cd} \mat{\pi(p-q)|\bar{b}\gamma_\mu
u} {B(p)}\mat{D^{(*)}|V_\mu}{0}~~~. \eeq
Using the standard parameterization \cite{BSW}, one obtains the ratio $r$
defined in Eq.~(\ref{def:r}):
\bea
r(D^* \pi) &=& \frac{f_{D^*}\,F_1^{B\pi}(m_{D^*}^2)}
                    {f_{\pi}\,A_0^{B D^*}(m_{\pi}^2)}, \nonumber \\
r(D \pi) &=& \frac{f_D\,(m_B^2-m_{\pi}^2)\,F_0^{B\pi}(m_D^2)}
                  {f_{\pi}\,(m_B^2-m_D^2)\,F_0^{BD}(m_{\pi}^2)}~~~.
\eea
In Table I, we give the values of $r$ for ${\bar B}^0 \to D^{(*)}\,\pi$ decays
in several models.  In all cases, the models predict that $r$ is close to
unity, i.e.\ $R \sim 0.02$.
\begin{table}
\label{table:1}
\centerline {
\begin{tabular}{ccc}
\hline\hline
 & $r$ ($D^* \pi$) & $r$ ($D \pi$) \\ \hline
Light Front Model \cite{JCCH} & 0.81 & 0.72 \\
BSW II Model \cite{NRSX} & 1.33 & 1.11 \\
NS Model \cite{NS} & 0.88 & 0.72 \\
LCSR Model \cite{LCSR} & 1.01 & 0.87 \\
MS Model \cite{MS} & 0.92 & 0.82 \\
\hline\hline
\end{tabular} \vspace{8pt}
}
\caption{The ratio $r$ evaluated in various models.}
\end{table}

The error on $R$ can be estimated using the method described in the
beginning of this Section.  Suppose that the number of detected $B^+ \to
D^{(*)+} \pi^0$ decays is $N_2$ out of $N$ tagged $B^+$'s, while the number 
of
detected $\bo \to D^{(*)-} \pi^+$ decays is $N_1$ out of the same number $N$
of tagged $B^0$'s. Then, assuming equal charged and neutral $B$ production,
the value
\beq
R=\sqrt{2\frac{\tau_\bo}{\tau_{B^+}}\,\frac{N_2}{N_1}}
\eeq
has an uncertainty
\beq
\sigma(R)=\sqrt{2\frac{\tau_\bo}{\tau_{B^+}}} 
\frac{1}{2\sqrt{N_1}}\sqrt{1+\frac{N_2}{N_1}}
\approx \frac{1}{\sqrt{2N_1}}.
\eeq
Taking into account $B^- \to D^{(*)-} \pi^0$ and $\ob \to D^{(*)+} \pi^-$
decays increases statistics by a factor of $2$, leading to
$\sigma(R) = 1/\sqrt{2N_r}$, where $N_r=2N_1$ is the
number of $\bo \to D^{(*)-} \pi^+$ decays plus the number $\ob \to D^{(*)+}
\pi^-$ decays.  To make connection with the total number $N_B$ of produced $B
\bar B$ pairs, note that the number of tagged events is $N_r=\epsilon\, {\cal
B}^r N_B$.  Thus,
\beq
\label{errora2a1}
\sigma(R)=\frac{1}{\sqrt{2\epsilon {\cal B}^r N_B}}~~~.
\eeq
For $10^8$ produced $B \bar B$ pairs $\sigma(R)=0.17 \cdot 10^{-2}$,
i.e.\ less than $10\%$ of its value.  Thus, measurements of $B^+ \to D^{(*)+}
\pi^0$ decay rates provide the ratio of amplitudes with a high precision.
This information may be used in the time-integrated approach discussed in the
previous Section.  Now we can go a step further and estimate the uncertainty
in determination of $\sin(2 \beta + \gamma)\cos\delta$ and $\cos(2\beta+\gamma)
\sin\delta$.

In the following analysis, we will take $r=1$ (corresponding to $R
\simeq 0.02$) and use Eq.~(\ref{errora2a1}) to estimate the error on the ratio
$R$.

\section{Uncertainties in $\sin(2\beta+\gamma)\cos\delta$ and 
$\cos(2\beta+\gamma)\sin\delta$ with perfect time resolution
and no mistagging}

The uncertainties in the ratios $f_2$ and $f_3$ [see
Eqs.~(\ref{eqn:f2}) and~(\ref{eqn:f3})] are
\beq
\sigma(f_2) \approx \sigma(f_3) \approx \frac{1}{\sqrt{N(t_0)}} = 
\sqrt{\frac{e^{\Gamma t_0}}{\epsilon ({\cal B}^r+{\cal B}^w) N_B}} =
\sqrt{\frac{e^{\Gamma t_0}}{\epsilon (1+k)\,{\cal B}^r N_B}}.
\eeq
Eqs.~(\ref{eqn:SC}) and~(\ref{eqn:CS}) allow an estimate of the values of 
$f_2$ and  $f_3$: $f_2 \approx 2R\,b(t_0)\,SC$, $f_3 \approx 2R\,b(t_0)\,CS$.
Now that Eq.~(\ref{errora2a1}) provides the error in $R$, we can calculate 
the uncertainties in $SC$ and $CS$:
\beq
\sigma(SC)  \approx  \frac{1}{2b(t_0)}\sqrt{\frac{f_2^2(t_0)}{R^4} 
\sigma^2(R) + \frac{\sigma^2(f_2)}{R^2}}
\leq  \frac{1} {2b(t_0) \,R} 
\sqrt{\frac{2(1+k)\,b^2(t_0)+e^{\Gamma t_0}}{\epsilon (1+k)\,{\cal B}^r 
N^B}},
\eeq
\beq
\sigma(CS)  \approx  \frac{1}{2b(t_0)}\sqrt{\frac{f_3^2(t_0)}{R^4} 
\sigma^2(R) + \frac{\sigma^2(f_3)}{R^2}}
\approx  \frac{\sigma(f_3)}{2b(t_0)\,R} 
\approx  \frac{1} 
{2b(t_0)\,R} \sqrt{\frac{e^{\Gamma t_0}}{\epsilon (1+k)\,
{\cal B}^r N^B}}.
\eeq
Finally, one
can calculate the number of $B \bar B$ pairs needed to get any particular
precision $\sigma_0(SC)$:
\begin{figure}
\centerline{\epsfysize = 3.6 in \epsffile{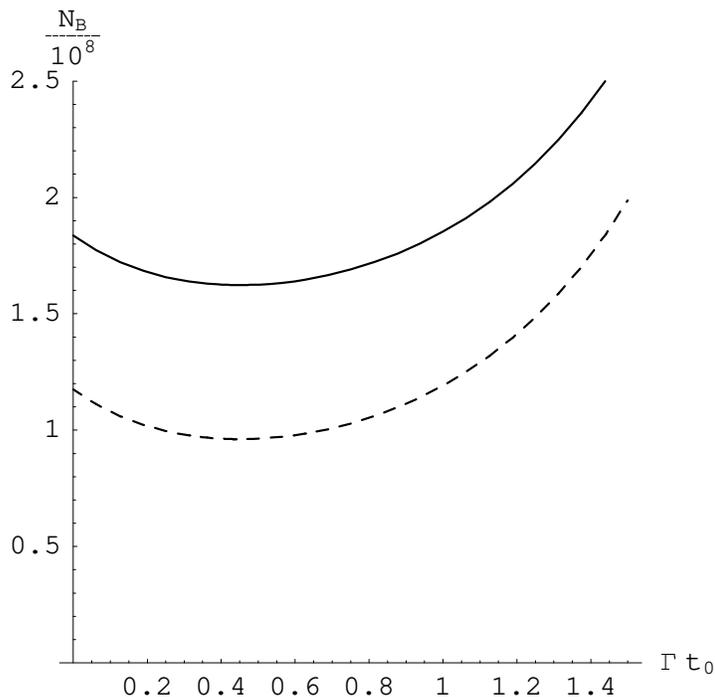}} 
\caption{Number of produced $B \bar B$ pairs needed to achieve an uncertainty
of $0.1$ in measurements of  $\sin(2\beta+\gamma)\cos\delta$ (solid line) and
$\cos(2\beta+\gamma)\sin\delta$ (dashed line) vs.\ minimum vertex separation
$t_0$. Perfect time resolution is assumed.}
\end{figure}
\beq
N_B \approx \frac{2(1+k)\,b^2(t_0)+e^{\Gamma t_0}}{4 \epsilon (1+k)\, 
b^2(t_0)
R^2 {\cal B}^r \sigma_0^2(SC)},
\eeq
or $\sigma_0(CS)$:
\beq
N_B \approx \frac{e^{\Gamma t_0}}{4 \epsilon (1+k)\, b^2(t_0) R^2 {\cal B}^r
\sigma_0^2(CS)}.
\eeq
As seen from the figure, these two quantities have the same minimum location
because they only differ by a constant independent of $t_0$.
Here we have assumed that $f_3(t_0)$ is proportional to
$\cos(2\beta+\gamma)\sin\delta$, which is expected to be small, and that $SC$
is close to 1.  However, the neglected $SC$ and $CS$ dependence can be readily
put back if necessary, and one finds that the position of the minima would
remain the same, independent of the values of $SC$ and $CS$, for both curves.

Fig.~2 shows the $N_B$ dependence on $t_0$ according to the above two
equations.  The curves were calculated under the assumption that one needs to
get $\sigma_0=0.1$. We found out that this precision level is sufficient to
determine $\sin(2\beta+\gamma)$ with an uncertainty of $0.05$ (Section~VII).
The optimal conditions for both measurements are achieved if one only takes
into account decays with vertex separation greater than $\sim 0.45/\Gamma$.
That one needs fewer $B {\bar B}$ pairs to reach the same precision for
$\cos(2\beta+\gamma)\sin\delta$ as indicated in Fig.~2 reflects our previous
assumption of small $\cos(2\beta+\gamma)\sin\delta$.  Thus, the minimum
uncertainties one can obtain if $N_B$ $B \bar B$ pairs are available are
\beq
\sigma_{min}(SC) \simeq 0.1\sqrt{\frac{1.62\cdot10^8}{N_B}},
\eeq
\beq
\sigma_{min}(CS) \simeq 0.1\sqrt{\frac{0.96\cdot10^8}{N_B}}.
\eeq 
Now we shall check how these formulae change if we take into account finite 
time resolution and realistic mistagging probabilities.

\section{Finite time resolution; mistagging}

Measurements of the decay numbers are smeared by finite resolution of vertex 
separation. For simplicity we shall assume a single Gaussian resolution 
function. The observed decay numbers are given by 
Eqs.~(\ref{eqn:nr})$-$(\ref{eqn:nw}) convoluted with the resolution function
\beq
R(t_0) \equiv \int_{-\infty}^{+\infty} d\mu \frac{1}{\sqrt{2\pi}\sigma}\, 
e^{-(\mu-t_0)^2/2\sigma^2}.
\eeq
For example,
\beq
N^r_\pm(t_0) = N_0\,R(t_0) \otimes N^r_\pm(\mu), 
\eeq
and similar convoluted relations for $N^w_\pm(t_0)$, $\overline{N^r_\pm}(t_0)$,
and $\overline{N^w_\pm}(t_0)$. Here $N_0$ is a normalization factor and
$\sigma$ is the resolution of time separation between vertices. For the BaBar
detector the average resolution of space separation between vertices is
$180\, \mu \mbox{m}$ \cite{sin2beta} while the average separation is
$\beta\gamma c \tau_{B^0}=260\, \mu \mbox{m}$, implying $\sigma\Gamma = 180/260
\cong 0.69$.

The algebraic sums of decay numbers that enter Eqs.~(\ref{eqn:f2})
and~(\ref{eqn:f3}) have to be modified correspondingly. For example,
Eq.~(\ref{eqn:totalN}) becomes
\bea
N(t_0)
& = & 2\frac{N_0}{\sqrt{2}\sigma}\,\int_{-\infty}^{+\infty} d\mu \,
\Phi'\left(\frac{\mu-t_0}{\sqrt{2}\sigma}\right) \, \int_\mu^{+\infty}
dt'\,e^{-\Gamma |t'|} \left[ A_+(t')+A_-(t') \right] \nonumber \\
& = & 2\frac{N_0}{\Gamma}\,\int_{-\infty}^{+\infty} d\tilde t\,\, e^{-|\tilde
t|} \left[ A_+(\tilde t/\Gamma)+A_-(\tilde t/\Gamma) \right]
\nonumber \\
& & \ {}+2\frac{N_0}{\Gamma}\,\int_{-\infty}^{+\infty} d\tilde \mu\,\,\,
\Phi\left(\frac{\tilde \mu-\Gamma t_0}{\sqrt{2}\sigma\Gamma}\right)
e^{-|\tilde \mu|} \left[ A_+(\tilde \mu/\Gamma)+A_-(\tilde \mu/\Gamma)
\right]
\nonumber \\
& = & 4\frac{N_0}{\Gamma}\,\left(1+\frac{A^2_2}{A^2_1}\right)(J_1+J_2),
\label{eqn:sumr'}
\eea
Similarly, one obtains
\bea
(N^r_- + N^w_- + \overline{N^r_+} + \overline{N^w_+}) -
(N^r_+ + N^w_+ + \overline{N^r_-} + \overline{N^w_-})
&=& 8\frac{N_0}{\Gamma}\,R\,J_3\,SC, \\
(N^r_+ + N^w_- + \overline{N^r_+} + \overline{N^w_-}) -
(N^r_- + N^w_+ + \overline{N^r_-} + \overline{N^w_+})
&=& 8\frac{N_0}{\Gamma}\,R\,J_3\,CS.
\eea
In the above equations, $\Phi(x) \equiv (2/\sqrt{\pi})\int_0^x e^{-z^2}dz$ is
the error function and
\bea
J_1 &\equiv& \int_{-\infty}^{+\infty} e^{-|\tilde t|} \,\,d\tilde t = 2,
\label{eqn:j'1} \\
J_2 &\equiv& \int_{-\infty}^{+\infty} d\tilde \mu\,\,\,\Phi\left(\frac{\tilde
\mu-\Gamma t_0}{\sqrt{2}\sigma\Gamma}\right)
e^{-|\tilde \mu|}, \\
J_3 &\equiv& \int_{-\infty}^{+\infty} d\tilde \mu\,\,\,\Phi\left(\frac{\tilde
\mu-\Gamma t_0}{\sqrt{2}\sigma\Gamma}\right)
e^{-|\tilde \mu|} \sin x_d \tilde \mu,
\label{eqn:j'3}
\eea
The last two integrals have been numerically evaluated for different values
of $t_0$ in the range from $0$ to $1.5/\Gamma$. Now $SC$ and $CS$ can be
rewritten in terms of the ratios $f_2$ and $f_3$ as
\beq
SC,CS = 
\frac{J_1+J_2}{2J_3}\, \frac{1+R^2}{R}\,f_{2,3}(t_0).
\eeq

Next, we will take into account the mistagging factor. Mistagging refers to
the cases where a decay ($B^0 \to$~tag, $\ob \to D^{*-} \pi^+$) was
incorrectly identified as ($\ob \to$~tag, $B^0 \to D^{*-} \pi^+$), and vice
versa. Thus, one sees that decays labelled as $B^0 \to D^{*-} \pi^+$
(``right-sign" decays) actually contain some $\ob \to D^{*-} \pi^+$
(``wrong-sign" decays) events. As a result, experimental measurements only
provide decay numbers smeared by the mistagging effect.  For instance, the
numbers of apparent right-sign events are
\beq
N^{r'}_\pm(t_0) = (1-w)N^r_\pm(t_0) + w \overline{N^w_\pm}(t_0),
\eeq
where $w$ is the mistagging probability. For the BaBar detector the tagging
efficiency is $\epsilon=\sum \epsilon_i = 0.684 \pm 0.007$ while the
effective tagging efficiency is $Q = \sum \epsilon_i (1-2w_i)^2=0.261 \pm
0.012$~\cite{sin2beta}. For our purposes we will simplify calculations by
assuming the single tagging option with $\epsilon=0.684$ and
$Q=\epsilon(1-2w)^2=0.261$. Thus, the effective mistagging probability is
$w=0.191$. 

Note that the sum of all smeared decay numbers is still equal to $N$, the sum
of all physical decay numbers.
One can show that the ratios $f'_2$ and $f'_3$ composed of smeared decay 
numbers are related to $f_2$ and $f_3$ by $f'_{2,3} = (1-2w)f_{2,3}$.
Thus, experimental measurements of smeared decay numbers allow
the direct calculations of $SC$ and $CS$:
\beq
{SC,CS} = \frac{1}{1-2w}\,\frac{J_1+J_2}{2J_3}\,
\frac{1+R^2}{R}\,f'_{2,3}(t_0).
\eeq
Assuming that experimental uncertainties are
$\sigma[N^{r'}_\pm(t_0)]=\sqrt{N^{r'}_\pm(t_0)}$,
$\sigma[N^{w'}_\pm(t_0)]=\sqrt{N^{w'}_\pm(t_0)}$, etc., we can estimate the
uncertainties in $f'_2$ and $f'_3$ measurements to be
\beq
\sigma(f'_2) \approx \sigma(f'_3) \approx \frac{1}{\sqrt{N(t_0)}} =
\sqrt{\frac{e^{\Gamma t_0}}{\epsilon (1+k)\,{\cal B}^r N_B}}.
\eeq
The uncertainties in $SC$ and $CS$ measurements are
\beq
\sigma(SC) \leq  \frac{1}{1-2w}\,\frac{J_1+J_2}{2J_3}\,\frac{1}{R}\,
\sqrt{\frac{2(1+k)\,[J_3/(J_1+J_2)]^2\,(1-2w)^2+e^{\Gamma t_0}}{\epsilon
(1+k)\,{\cal B}^r N^B}},
\eeq
\beq 
\sigma(CS) \approx \frac{1}{1-2w}\,\frac{J_1+J_2}{2J_3}\,\frac{1}{R}\,
\sqrt{\frac{e^{\Gamma t_0}}{\epsilon (1+k)\,{\cal B}^r N^B}}.
\eeq
We assumed a small $CS$ in deriving the second equation. Finally, one can
calculate the number of $B \bar B$ pairs needed to get any particular
precision $\sigma_0(SC)$:
\beq
N_B \approx 
\frac{2(1-2w)^2(1+k)\,J_3^2+e^{\Gamma t_0}\,(J_1+J_2)^2} {4\epsilon (1-2w)^2
(1+k)\,J_3^2\,R^2 {\cal B}^r \sigma_0^2(SC)},
\eeq
or $\sigma_0(CS)$:
\beq
N_B \approx
\frac{e^{\Gamma t_0}\,(J_1+J_2)^2} {4\epsilon (1-2w)^2 (1+k)\,J_3^2\,R^2
{\cal B}^r \sigma_0^2(CS)}.
\eeq
As in the previous Section,
the position of the minima is the same for both curves and is independent of
the values of $SC$ and $CS$.

\begin{figure}
\centerline{\epsfysize = 3.6 in \epsffile{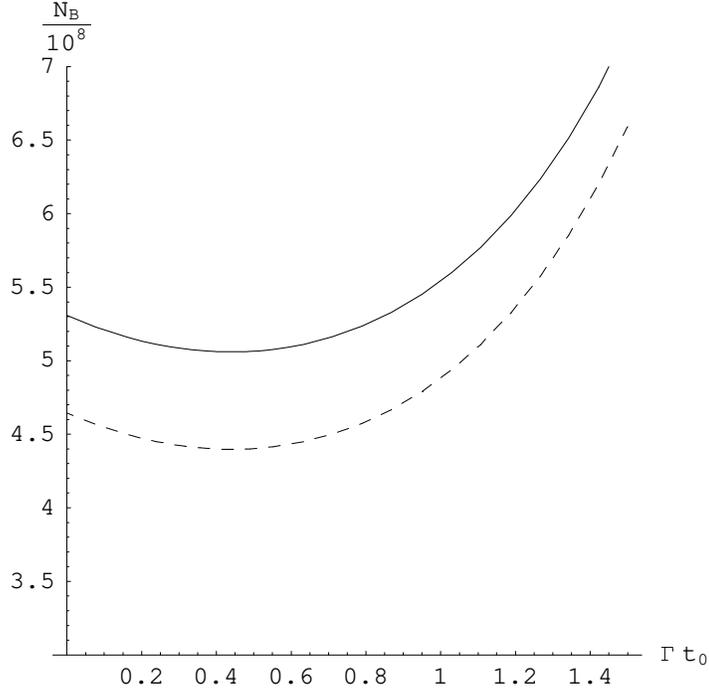}} 
\caption{Number of produced $B \bar B$ pairs needed to achieve the
uncertainty of $0.1$ in measurements of  $\sin(2\beta+\gamma)\cos\delta$
(solid line) and $\cos(2\beta+\gamma)\sin\delta$ (dashed line) vs.\ minimum
vertex separation $t_0$.}
\end{figure}
Fig.~3 shows the $N_B$ dependence on $t_0$. The curves were calculated under
the assumption that one needs to get $\sigma_0=0.1$. The optimal conditions
for measurements are achieved if one only takes into account decays with
vertex separation greater than $0.44/\Gamma$. Then
\beq
\sigma_{min}(SC) \simeq 0.1\sqrt{\frac{5.06\cdot10^8}{N_B}},
\label{eqn:sigSC}
\eeq
\beq
\sigma_{min}(CS) \simeq 0.1\sqrt{\frac{4.40\cdot10^8}{N_B}}.
\label{eqn:sigCS}
\eeq 

If BaBar is able to improve its performance to the level quoted
in~\cite{Bas2b}, i.e.\ $\sigma(\Delta z)=110\ \mu \mbox{m}$, $\epsilon=0.767$
and $Q=0.279$, then the required minimum number of $B {\bar B}$ pairs reduces
by a factor of $\sim 1.4$ for both $SC$ and $CS$ measurements. Besides, the
position of the minima is shifted to a slightly larger value ($t_0 \sim
0.53/\Gamma$) of vertex separation.
  
\section{Extraction of $\sin(2\beta+\gamma)$ and $\cos\delta$}

\begin{figure}
\centerline{\epsfysize = 7.3 in \epsffile{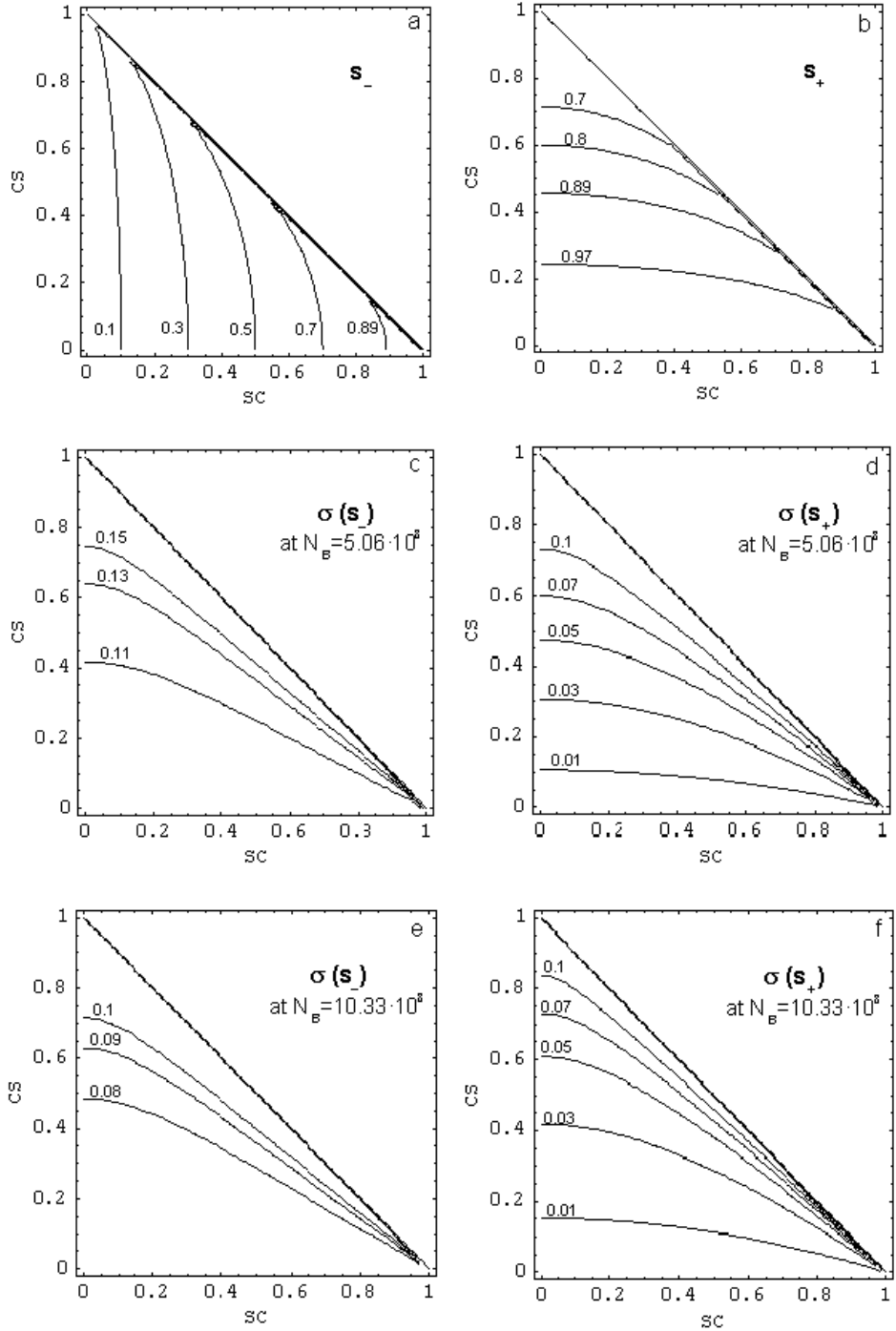}}
\caption{Contours of $s_-$, $s_+$, and their uncertainties $\sigma(s_-)$ and
$\sigma(s_+)$ in the ($SC$,$CS$) plane. Only the first quadrant of the plane
is shown. The plots in other quadrants are symmetric to those in the first
one since the plotted quantities only depend on the {\it absolute} values of
$SC$ and $CS$. The blank triangle above the line $SC+CS=1$ denotes the
forbidden region on the plane: $SC+CS=\sin(2\beta+\gamma+\delta)$ should
always be smaller or equal to $1$. When $CS=0\ $ $\sigma(s_-)$ achieves its
smallest values: $0.1$ and $0.07$ for plots (c) and (e) respectively.}
\end{figure}

If one measures $\sin(2\beta+\gamma)\cos\delta$ and
$\cos(2\beta+\gamma)\sin\delta$ values to be $SC$ and $CS$, then trigonometry
dictates the following values for $\sin^2(2\beta+\gamma)$ {\it and}
$\cos^2\delta$:
\beq
\sin^2(2\beta+\gamma),\cos^2\delta = s^2_{\pm} \equiv
\frac12\left(1+SC^2-CS^2\pm\sqrt{\lambda(1,SC^2,CS^2)}\right),
\label{eqn:ambiguity}
\eeq
where $\lambda(x,y,z) \equiv x^2+y^2+z^2-2xy-2yz-2zx$.  When one root
corresponds to $\sin^2(2\beta+\gamma)$, the other corresponds to
$\cos^2\delta$.
There is an ambiguity: Which is which? One cannot resolve it without making
additional assumptions.

If the value of $\sin(2\beta+\gamma)$ is assumed to be in agreement with the
indirect bounds (Fig.~1) then it should be larger then $0.89$.
However, indications that $\cos\delta$ is large, too~\cite{KS}, do not allow
an easy distinction between the two quantities. Figs.~4(a,b) show the
contours of  $s_-$ and $s_+$ values in the ($SC$,$CS$) plane. There is a big
region on the plane where $s_-<0.89$ while $s_+>0.89$. If the measured values
of $SC$ and $CS$ fall inside this region then $\sin(2\beta+\gamma)$ could
only be associated with $s_+$ and the ambiguity might be resolved. The
possibility of resolution also depends on the uncertainties $\sigma(s_-)$ and
$\sigma(s_+)$. Those are calculated from Eq.~(\ref{eqn:ambiguity}) with the
help of Eqs.~(\ref{eqn:sigSC})--(\ref{eqn:sigCS}). The contours of these
uncertainties are shown in Figs.~4(c,d,e,f) for two different numbers of $B
\bar B$ pairs. For example, if the number of produced $B \bar B$ pairs is
$5.06\cdot10^8$ and $(SC,CS)=(0.75,0.15)$, then we can calculate $s_+ = 0.97
\pm 0.04$ and
$s_- = 0.77 \pm 0.12$. In this case, $s_-$ does not take values that are
larger than $0.89$ within the $1\sigma$ level, and the solution favored for
consistency with Fig.~1 is $\sin^2(2\beta+\gamma)=s_+^2$,
$\cos^2\delta=s_-^2$. The 4-fold ambiguity in $2\beta+\gamma$ remains but
reduces to a 2-fold one when we take into account that only positive values
of $\sin(2\beta+\gamma)$ are consistent with indirect bounds.
One can see from Fig.~1 that if $2\beta+\gamma<\pi/2$ then
$\sin(2\beta+\gamma)$ should be larger than $0.97$. This fact might
completely resolve the ambiguity in favor of $\pi/2<2\beta+\gamma<\pi$ if
values larger than $0.97$ are measured to be  inconsistent with
$\sin(2\beta+\gamma)$ within the $1\sigma$ level.

Of course, one cannot exclude the possibility that $\sin(2\beta+\gamma)$ is
inconsistent with indirect bounds and is substantially smaller than $0.89$
while $|\cos\delta|$ is close to unity. In that case, one would make a wrong
assignment of $s_+$ and $s_-$ to $\sin(2\beta+\gamma)$ and $\cos\delta$,
respectively. Therefore, it is preferable to make other measurements of
$\sin(2\beta+\gamma)$ in decays like $B \to D^{(*)} \rho$ or $B \to D^{(*)}
a_1$ where strong phase might differ from $\delta$ in $B \to D^{(*)} \pi$
decays.

It is also worth noting that for the overwhelming part of the region where
$s_+>0.89$, the uncertainty in $\sin(2\beta+\gamma)$ is at most $0.05$ [cf.\
Figs.~4(b) and~4(d)]. Thus, for many values of $SC$ and $CS$ this method
allows a very precise determination of $\sin(2\beta+\gamma)$ and a good
measurement of the strong phase $\delta$.

Besides, the method can be used to detect deviations from the Standard Model.
If the measured values of $\sin(2\beta+\gamma)\cos\delta$ and
$\cos(2\beta+\gamma)\sin\delta$ fall into the upper left corner of the
($SC$,$CS$) plane, then both $s_-$ and $s_+$ would be inconsistent with the
$0.89-1.0$ range expected from the unitarity of the CKM matrix.
 
\section{Conclusions}

This paper has explored the optimal conditions for measurements of weak phase
angle $2\beta+\gamma$ and strong phase $\delta$ between Cabibbo-allowed and
doubly-Cabibbo-suppressed amplitudes in $B \to D^{(*)}\pi$ decays.
We have found that in the time-integrated approach it is advantageous to only
consider events with vertex separation greater than $t_0$ which is equal to
$0.44/\Gamma$ for the BaBar detection parameters. The loss in statistics is
outweighed by an increase in the integrated asymmetry.

Fig.~3 shows that production of approximately $5 \cdot 10^8$ $B \bar B$ pairs
is needed to reduce the uncertainty in determination of
$\sin(2\beta+\gamma)\cos\delta$ to $0.1$ in $B \to D^* \pi$ decays. A smaller
error on $\cos(2\beta+\gamma)\sin\delta$ will be achieved at the same time
if its value is small. $B \to D\pi$ decays have the advantage of a slightly
higher branching ratio but a setback in $D$ meson detection. The combination
of both types of decays might reduce the number of needed $B \bar B$ pairs
to $2.5\cdot 10^8$, an amount within the reach of both BaBar and BELLE in
the next few years. A time-dependent analysis \cite{BaBartd} does not lead to
any improvement with respect to this figure.

If the strong phase $\delta$ is not very close to $0$ or $\pi$, the ambiguity
between $\sin(2\beta+\gamma)$ and $\cos\delta$ can be resolved. This method
allows $\sin(2\beta+\gamma)$ to be determined with a precision of $0.05$ or
better.

\section*{Acknowledgments}

We thank I. Dunietz, M. Gronau, Z. Ligeti, Z. Luo, S. Prell and A. Soffer for
helpful discussions. This work was supported in part by the United States
Department of Energy through Grant Nos.\ DE-FG02-90ER-40560 and
W-31109-ENG-38.  C.-W. C. would like to thank the Summer Visitor Program
held by the Theory Department at Fermilab for their hospitality, and J. L.
R. thanks the Aspen Center for Physics, where part of this work was done.

\def \ajp#1#2#3{Am.\ J. Phys.\ {\bf#1}, #2 (#3)}
\def \apny#1#2#3{Ann.\ Phys.\ (N.Y.) {\bf#1}, #2 (#3)}
\def \app#1#2#3{Acta Phys.\ Polonica {\bf#1}, #2 (#3)}
\def \arnps#1#2#3{Ann.\ Rev.\ Nucl.\ Part.\ Sci.\ {\bf#1}, #2 (#3)}
\def \art{and references therein}
\def \cmts#1#2#3{Comments on Nucl.\ Part.\ Phys.\ {\bf#1}, #2 (#3)}
\def \cn{Collaboration}
\def \cp89{{\it CP Violation,} edited by C. Jarlskog (World Scientific,
Singapore, 1989)}
\def \efi{Enrico Fermi Institute Report No.\ }
\def \epjc#1#2#3{Eur.\ Phys.\ J. C {\bf#1}, #2 (#3)}
\def \f79{{\it Proceedings of the 1979 International Symposium on Lepton and
Photon Interactions at High Energies,} Fermilab, August 23-29, 1979, ed. by
T. B. W. Kirk and H. D. I. Abarbanel (Fermi National Accelerator Laboratory,
Batavia, IL, 1979}
\def \hb87{{\it Proceeding of the 1987 International Symposium on Lepton and
Photon Interactions at High Energies,} Hamburg, 1987, ed. by W. Bartel
and R. R\"uckl (Nucl.\ Phys.\ B, Proc.\ Suppl., vol.\ 3) (North-Holland,
Amsterdam, 1988)}
\def \ib{{\it ibid.}~}
\def \ibj#1#2#3{~{\bf#1}, #2 (#3)}
\def \ichep72{{\it Proceedings of the XVI International Conference on High
Energy Physics}, Chicago and Batavia, Illinois, Sept. 6 -- 13, 1972,
edited by J. D. Jackson, A. Roberts, and R. Donaldson (Fermilab, Batavia,
IL, 1972)}
\def \ijmpa#1#2#3{Int.\ J.\ Mod.\ Phys.\ A {\bf#1}, #2 (#3)}
\def \ite{{\it et al.}}
\def \jhep#1#2#3{JHEP {\bf#1}, #2 (#3)}
\def \jpb#1#2#3{J.\ Phys.\ B {\bf#1}, #2 (#3)}
\def \lg{{\it Proceedings of the XIXth International Symposium on
Lepton and Photon Interactions,} Stanford, California, August 9--14 1999,
edited by J. Jaros and M. Peskin (World Scientific, Singapore, 2000)}
\def \lkl87{{\it Selected Topics in Electroweak Interactions} (Proceedings of
the Second Lake Louise Institute on New Frontiers in Particle Physics, 15 --
21 February, 1987), edited by J. M. Cameron \ite~(World Scientific, Singapore,
1987)}
\def \kdvs#1#2#3{{Kong.\ Danske Vid.\ Selsk., Matt-fys.\ Medd.} {\bf #1},
No.\ #2 (#3)}
\def \ky85{{\it Proceedings of the International Symposium on Lepton and
Photon Interactions at High Energy,} Kyoto, Aug.~19-24, 1985, edited by M.
Konuma and K. Takahashi (Kyoto Univ., Kyoto, 1985)}
\def \mpla#1#2#3{Mod.\ Phys.\ Lett.\ A {\bf#1}, #2 (#3)}
\def \nat#1#2#3{Nature {\bf#1}, #2 (#3)}
\def \nc#1#2#3{Nuovo Cim.\ {\bf#1}, #2 (#3)}
\def \nima#1#2#3{Nucl.\ Instr.\ Meth. A {\bf#1}, #2 (#3)}
\def \np#1#2#3{Nucl.\ Phys.\ {\bf#1}, #2 (#3)}
\def \npbps#1#2#3{Nucl.\ Phys.\ B Proc.\ Suppl.\ {\bf#1}, #2 (#3)}
\def \os{XXX International Conference on High Energy Physics, Osaka, Japan,
July 27 -- August 2, 2000}
\def \PDG{Particle Data Group, D. E. Groom \ite, \epjc{15}{1}{2000}}
\def \pisma#1#2#3#4{Pis'ma Zh.\ Eksp.\ Teor.\ Fiz.\ {\bf#1}, #2 (#3) [JETP
Lett.\ {\bf#1}, #4 (#3)]}
\def \pl#1#2#3{Phys.\ Lett.\ {\bf#1}, #2 (#3)}
\def \pla#1#2#3{Phys.\ Lett.\ A {\bf#1}, #2 (#3)}
\def \plb#1#2#3{Phys.\ Lett.\ B {\bf#1}, #2 (#3)}
\def \pr#1#2#3{Phys.\ Rev.\ {\bf#1}, #2 (#3)}
\def \prc#1#2#3{Phys.\ Rev.\ C {\bf#1}, #2 (#3)}
\def \prd#1#2#3{Phys.\ Rev.\ D {\bf#1}, #2 (#3)}
\def \prl#1#2#3{Phys.\ Rev.\ Lett.\ {\bf#1}, #2 (#3)}
\def \prp#1#2#3{Phys.\ Rep.\ {\bf#1}, #2 (#3)}
\def \ptp#1#2#3{Prog.\ Theor.\ Phys.\ {\bf#1}, #2 (#3)}
\def \rmp#1#2#3{Rev.\ Mod.\ Phys.\ {\bf#1}, #2 (#3)}
\def \rp#1{~~~~~\ldots\ldots{\rm rp~}{#1}~~~~~}
\def \si90{25th International Conference on High Energy Physics, Singapore,
Aug. 2-8, 1990}
\def \slc87{{\it Proceedings of the Salt Lake City Meeting} (Division of
Particles and Fields, American Physical Society, Salt Lake City, Utah, 1987),
ed. by C. DeTar and J. S. Ball (World Scientific, Singapore, 1987)}
\def \slac89{{\it Proceedings of the XIVth International Symposium on
Lepton and Photon Interactions,} Stanford, California, 1989, edited by M.
Riordan (World Scientific, Singapore, 1990)}
\def \smass82{{\it Proceedings of the 1982 DPF Summer Study on Elementary
Particle Physics and Future Facilities}, Snowmass, Colorado, edited by R.
Donaldson, R. Gustafson, and F. Paige (World Scientific, Singapore, 1982)}
\def \smass90{{\it Research Directions for the Decade} (Proceedings of the
1990 Summer Study on High Energy Physics, June 25--July 13, Snowmass, 
Colorado),
edited by E. L. Berger (World Scientific, Singapore, 1992)}
\def \tasi{{\it Testing the Standard Model} (Proceedings of the 1990
Theoretical Advanced Study Institute in Elementary Particle Physics, Boulder,
Colorado, 3--27 June, 1990), edited by M. Cveti\v{c} and P. Langacker
(World Scientific, Singapore, 1991)}
\def \yaf#1#2#3#4{Yad.\ Fiz.\ {\bf#1}, #2 (#3) [Sov.\ J.\ Nucl.\ Phys.\
{\bf #1}, #4 (#3)]}
\def \zhetf#1#2#3#4#5#6{Zh.\ Eksp.\ Teor.\ Fiz.\ {\bf #1}, #2 (#3) [Sov.\
Phys.\ - JETP {\bf #4}, #5 (#6)]}
\def \zpc#1#2#3{Zeit.\ Phys.\ C {\bf#1}, #2 (#3)}
\def \zpd#1#2#3{Zeit.\ Phys.\ D {\bf#1}, #2 (#3)}

\end{document}